\documentclass[iop,apjl]{emulateapj}
\usepackage{epsfig}
\usepackage{amsmath, amsthm, amssymb}
\usepackage[plainpages=false, colorlinks=true, anchorcolor=blue, linkcolor=blue, citecolor=blue, bookmarks=false, urlcolor=blue]
{hyperref}
\usepackage{color}
\usepackage{microtype}
\usepackage{float}
\usepackage{verbatim}
\usepackage{wrapfig}
\usepackage[caption = false]{subfig}
\usepackage{multirow}
\usepackage{hhline}
\usepackage{booktabs}
\usepackage{array}
\usepackage{tabularx,ragged2e,booktabs}
\usepackage{appendix}

\newcommand{\CatalinaTotalFlagged}{68,076}

\newcommand{\CatalinaGoldFlagged}{39,600}
\newcommand{\CatalinaSilverFlagged}{6,492}
\newcommand{\CatalinaBronzeFlagged}{19,134}
\newcommand{\CatalinaGsigmaFalsePositivesWithBPandRP}{65,226}

\newcommand{\RRLyraeTotal}{107,418 }

\newcommand{\RRLyraeFlagged}{75,911 }
\newcommand{\RRLyraeGFlagged}{107,153 }
\newcommand{\RRLyraeGoldFlagged}{26,211 }
\newcommand{\RRLyraeSilverFlagged}{11,755 }
\newcommand{\RRLyraeBronzeFlagged}{37,945 }

\newcommand{\CepheidTotal}{8,465 }

\newcommand{\CepheidFlagged}{7,661 }

\newcommand{\CepheidGoldFlagged}{676 }
\newcommand{\CepheidSilverFlagged}{121 }
\newcommand{\CepheidBronzeFlagged}{6,864 }

\newcommand{\ZwickyTotal}{556,521 }
\newcommand{\ZwickyFlagged}{361,850 }
\newcommand{\ZwickyGFlagged}{524,634 }
\newcommand{\ZwickyGoldFlagged}{83,246 }
\newcommand{\ZwickySilverFlagged}{57,726 }
\newcommand{\ZwickyBronzeFlagged}{220,878 }

\newcommand{\StripeBronzeGalaxiesRemoved}{16,979}

\newcommand{\StripeBronzeNEWflagged}{17,777}

\newcommand{\StripeGoldGalaxiesRemoved}{820}

\newcommand{\StripeSilverGalaxiesRemoved}{719}

\newcommand{\GalaxiesRemoved}{4,734,587 }
\newcommand{\NeighborsRemoved}{42,670,766}

\newcommand{\fourGsig}{47,256,102 }
\newcommand{\fiveGsig}{40,385,829 }
\newcommand{\initialCatalogList}{56,704,871}
\newcommand{\CatalogNeighborsRemoved}{14,034,105}

\newcommand{\FinalTotalStarsFlagged}{9,299,518}
\newcommand{\FinalGoldFlagged}{269,576}
\newcommand{\FinalSilverFlagged}{483,006}
\newcommand{\FinalBronzeFlagged}{8,546,936}

\newcommand{\slugcom}{Accepted to ApJ}% on \today}
\slugcomment{\slugcom}

\shorttitle{Variables in Gaia DR2}
\shortauthors{Andrew et al.}

\usepackage{graphicx}
\usepackage{dcolumn}
\usepackage{bm}

\begin{document}

\title{\textbf{Identifying Candidate Optical Variables Using \textit{Gaia} Data Release 2}}

\author{Shion Andrew\altaffilmark{1},
Samuel J. Swihart\altaffilmark{2},
Jay Strader\altaffilmark{2}}

\affil{
    \altaffilmark{1}{Department of Physics, Harvey Mudd College, Claremont, CA 91711, USA}\\
    \altaffilmark{2}{Center for Data Intensive and Time Domain Astronomy, Department of Physics and Astronomy,\\Michigan State University, East Lansing, MI 48824, USA\\}
}

\begin{abstract}
\emph{Gaia} is undertaking a deep synoptic survey of the Galaxy, but photometry from individual epochs has, as of yet, only been released for a minimal number of sources. We show that it is possible to identify variable stars in \emph{Gaia} Data Release 2 by selecting stars with unexpectedly large photometric uncertainties given their brightness and number of observations. By comparing our results to existing catalogs of variables, we show that information on the amplitude of variability is also implicitly present in the \emph{Gaia} photometric uncertainties. We present a catalog of about 9.3 million candidate variable stars, and discuss its limitations and prospects for future tests and extensions.

\end{abstract}

\pacs{Valid PACS appear here}
\maketitle

\section{\label{sec:intro}Introduction}
A central theme of modern astrophysics is the shift from static catalogs of celestial sources to an increasing focus on variables and transients. One reason variability is of interest is because a complete inventory of variable stars down to almost any magnitude limit would provide important insights into stellar evolution and Galactic structure. For instance, past discoveries of variable stars have helped with the determination of cosmological parameters (e.g., \citealt{Freedman01,Riess16}), improving the distance scale (e.g., \citealt{Pietrzy13,Thompson01}), and calculating the ages of the oldest stars (e.g., \citealt{Soszy15}).

In our own Galaxy, recent wide-field surveys that have produced catalogs of variable stars include, e.g., the Catalina Real-Time Transient Survey (CRTS; \citealt{Drake14}), the All-Sky Automated Survey for Supernovae (ASAS-SN; \citealt{Jayasinghe18}), the Zwicky Transient Facility time-domain survey \citep{Bellm19}, and the \emph{Transiting Exoplanets Survey Satellite} (\emph{TESS}; \citealt{Ricker15}), among others. These variable catalogs are distinguished by their overarching science goals and therefore have different selections functions. For example, CRTS avoids the Galactic Plane and the poles, while ASAS-SN is limited to just the brighter sources with $g \lesssim 18$. 

A true deep, all-sky variability survey is ongoing with the astrometric \emph{Gaia} mission. However, the most recent \textit{Gaia} data release (Data Release Two, hereafter DR2\footnote{https://www.cosmos.esa.int/web/gaia/dr2}) only presented variability information for approximately 550,000 sources \citep{Holl18}, with per-epoch photometry for the full set of sources expected to follow in future releases.

In this paper we show that incomplete---but still quite valuable---information on which \emph{Gaia} sources are variable is implicitly present in DR2.

In Section~\ref{sec:background} we provide a short summary of \textit{Gaia} photometry and assess its advantages over other surveys in the search for optical variables. We summarize in Section~\ref{sec:data+methods} the algorithm we developed to identify variable stars using uncertainties in flux as measured by \textit{Gaia}. In Section~\ref{sec:Results} we test the algorithm against existing catalogs of variables and present our final catalog of candidate variable stars.

\section{Background: \emph{Gaia} Photometry}
\label{sec:background}
Since 2014, \textit{Gaia} has been collecting photometric and astrometric observations of the entire sky. \textit{Gaia} DR2, based on data taken during the first 22 months of the mission \citep{GaiaCollaboration18}, provides broadband $G$ photometry for approximately 1.7 billion sources. \textit{Gaia} DR2 also presents photometry in blue and red ($G_{BP}$ and $G_{RP}$) bands for a subset of sources from separate photometers; these data will eventually also give low-resolution spectra. \textit{Gaia} photometry is calibrated to a consistent and homogeneous photometric system \citep{Lindegren18}, and its location above the Earth's atmosphere offers similar advantages to \emph{TESS}, trading off the higher time resolution of \emph{TESS} for better depth and spatial resolution. This makes \emph{Gaia} especially well-suited for dense regions of the sky, with a pixel size of $\sim 0.06 \times 0.18\arcsec$, compared to \emph{TESS} or ASAS-SN which have typical pixel sizes of 21\arcsec and 8\arcsec, respectively.

\emph{Gaia} continually scans the sky by both spinning and slowly precessing. This scanning law gives complete, but not uniform, coverage of the sky. In addition, each transit of the focal plane by a source provides not one, but nine independent $G$ measurements from different CCDs, though other parameters such as the source density also affect the number of observations made of each source. Overall, nearly all sources listed in the photometric catalog have many observations, and the listed photometry is a weighted average of the individual measurements \citep{Evans18,Riello18}.

Individual epoch photometry has been published for a minimal fraction of the full \emph{Gaia} dataset, principally objects with many high-quality data points belonging to well-studied classes of variables such as RR Lyraes, Cepheids, and long-period variables \citep{Holl18}. Short-timescale variability has been explored for a yet smaller pilot sample of sources \citep{Roelens18}. These works have proven the usefulness of \emph{Gaia} photometry for studying stellar variability, but these catalogs are far from---and make no representation of being---a complete sample of variables. Indeed, the non-uniform \emph{Gaia} DR2 sky coverage, combined with a minimum required number of focal plane transits in \citet{Holl18}, means that some areas of the sky have literally zero identified variables in that paper.

Here we show that even without per epoch photometry, it is possible to accurately identify a large sample of variable stars using the published \emph{Gaia} DR2 catalog. In particular, we show that variable stars can be selected by targeting stars with unexpectedly high photometric uncertainties given their brightness and number of observations.

\begin{figure}[t!]
\includegraphics[width=\linewidth]{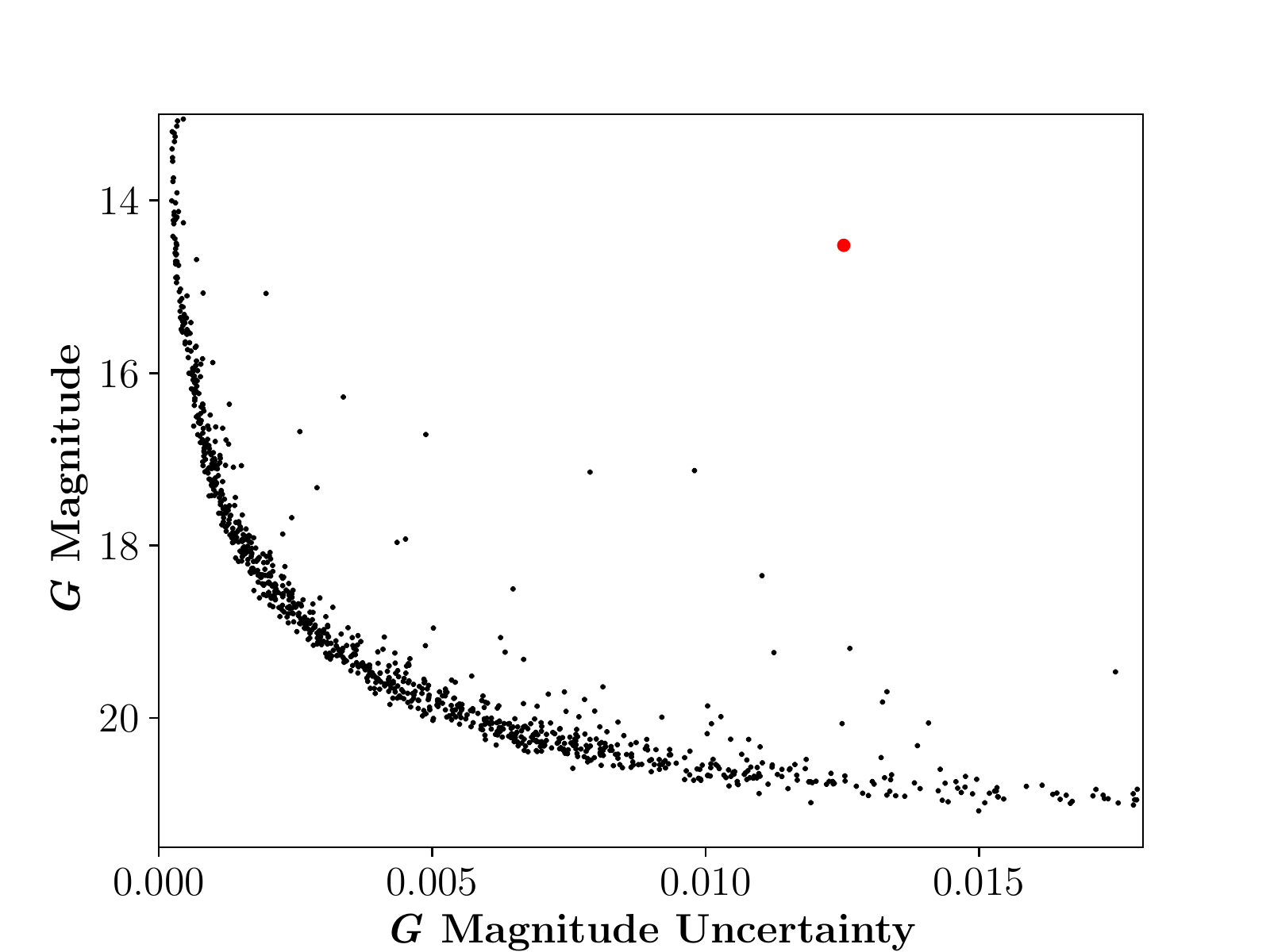}
\caption{A magnitude versus magnitude uncertainty plot illustrating the characteristic baseline curve along which most non-variable stars lie. This plot contains 1000 stars, obtained using a cone search of radius 0.3 degrees centered on the known RR Lyrae variable TY Hyi at (R.A., Dec.) of (38.2627379, --71.0328164). This variable star, colored in red, does not fall on the baseline curve, but instead has a noticeably larger magnitude uncertainty than other stars of comparable magnitude.}
\label{fig:sampleVariable}
\end{figure}

\section{Data and Methods}
\label{sec:data+methods}

\subsection{Photometric uncertainty encodes variability}
The essence of our method is shown in Figure~\ref{fig:sampleVariable}, which plots the \emph{Gaia} $G$ versus the $G$ magnitude uncertainty for 1000 stars in a small patch of the sky (due to our signal-to-noise cut described below, we assume symmetric magnitude uncertainties). The expected shot-noise curve, with larger uncertainties for fainter stars, is evident in Figure~\ref{fig:sampleVariable} for the bulk of the stars (see also \citealt{Evans18}). We refer to this distribution as the ``baseline curve''.

This field was not chosen at random: it is centered on a well-known RR Lyrae variable star, TY Hyi ($G=14.3$). Figure~\ref{fig:sampleVariable} shows that this star has a much larger $G$ uncertainty than stars of similar brightness on the baseline curve, consistent with the idea that intrinsic variability can increase the magnitude uncertainty of a star.

\begin{figure}[t]
\includegraphics[width=\linewidth]{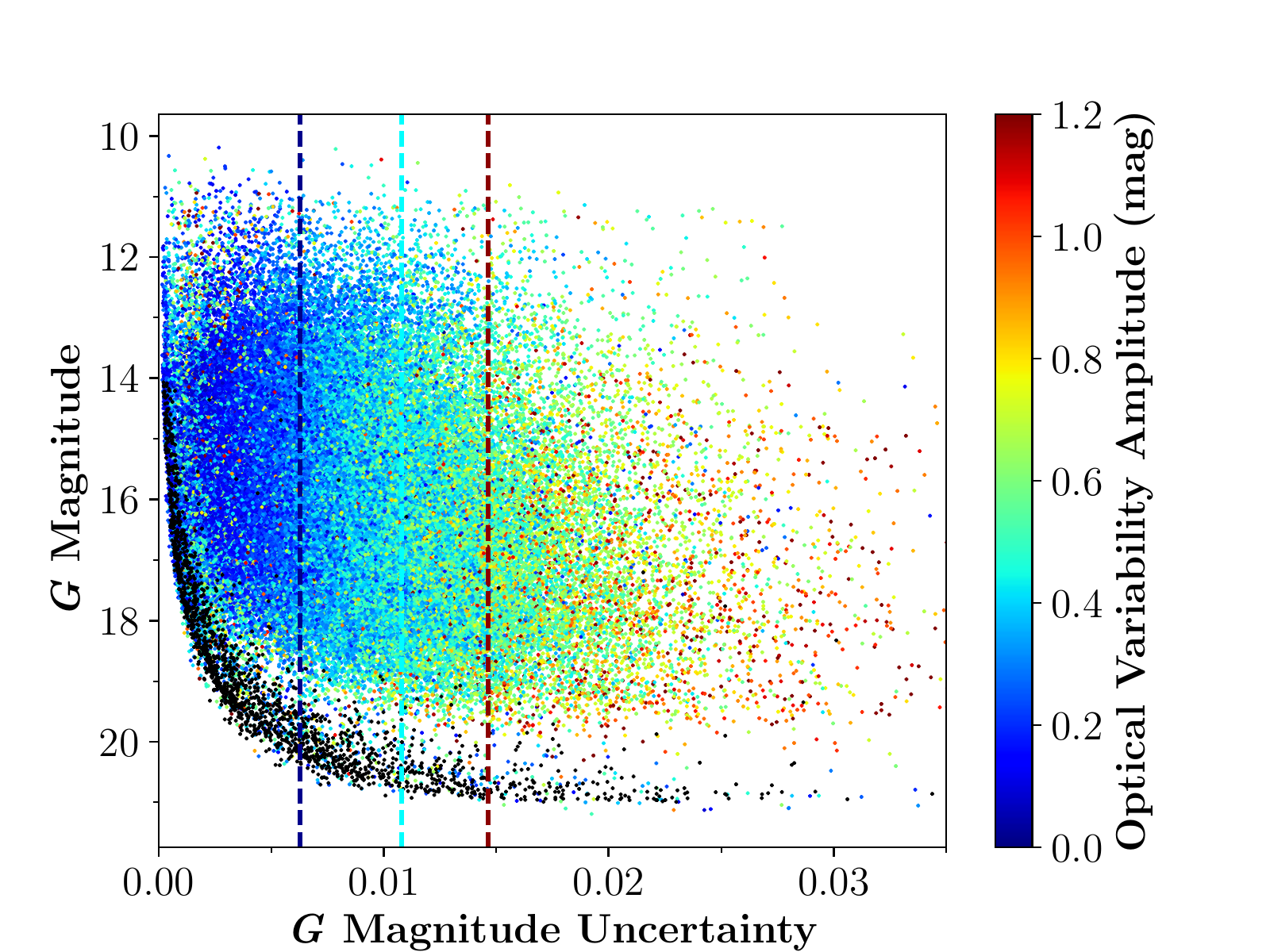}
\caption{$G$ vs. $G$ uncertainty for CRTS variables with periods less than 10 days \citep{Drake14,Drake17}, colored by their optical variability amplitude. The black points are a random sample of 2000 stars, illustrating a baseline curve for non-variable stars. 
The dashed lines are the mean magnitude uncertainty of variables, in three bins from 0.0 to 1.2 mag in variability amplitude.} The $G$ magnitude uncertainty increases with amplitude, implying that stars with greater variability lie further off the baseline curve.
\label{fig:CatalinaVariables}
\end{figure}

In Figure~\ref{fig:CatalinaVariables} we show a similar plot of $G$ versus magnitude uncertainty but now for all 70,680 published CRTS variable stars with photometric periods $< 10$ d, a conservative limit chosen to ensure that variability would be sampled by the \emph{Gaia} data. It is evident that few of these stars sit on the baseline curve, but instead, nearly all lie \emph{above} the curve, with a larger magnitude uncertainty than expected for other (non-variable) stars of similar brightness.

More evidence that this effect is due to intrinsic variability comes from considering the amplitude of the variability. The points in Figure \ref{fig:CatalinaVariables} are color-coded by variability amplitude. Stars with larger variability amplitudes deviate more from the baseline curve at a fixed magnitude, consistent with the idea that the magnitude uncertainty is correlated with variability.

\subsection{\label{sec:Baseline} Number of Observations and Calculating the Baseline Curve}

\subsubsection{$G$ Photometry}

Since the \emph{Gaia} sky coverage is non-uniform, and the $G$ magnitude uncertainty also varies with the number of times a source has been observed \citep{Evans18}, different samples of stars across the sky will produce different baseline curves. Hence a simple magnitude-based statistic for selecting variables will not work---the number of observations must also be considered.

\begin{figure}[t]
\includegraphics[width=\linewidth]{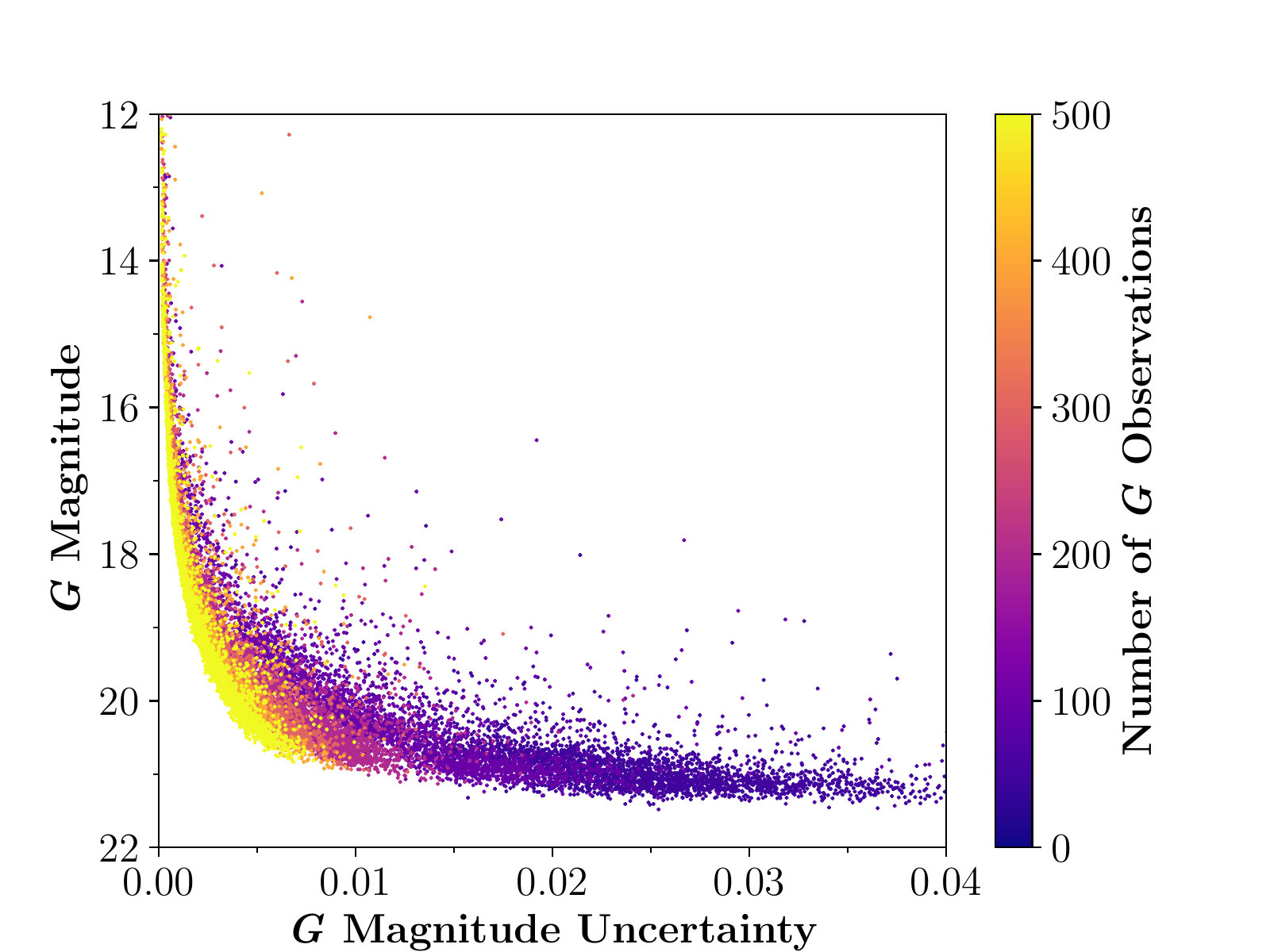}
\caption{$G$ mag vs. mag uncertainty plot, with the color representing the number of observations contributing to the \textit{Gaia} $G$ photometry. As expected, sources with more observations have smaller mean uncertainties and hence different baseline curves.}
\label{fig:baselineCurves}
\end{figure}

Figure~\ref{fig:baselineCurves} shows that, on average, the magnitude uncertainties scale as expected with the number of observations: at fixed brightness, the uncertainty decreases as the number of observations increases.
To model this, we binned stars by the number of $G$ observations, $N_{obs,G}$ , from 50 to 750 observations in intervals of 50. We further binned stars by magnitude, ranging from $G = 14$ to $G = 19.5$ in intervals of 0.1 mag. The bright limit was chosen to avoid an unusual feature in the magnitude uncertainty of stars in the range of about $G = 12.5-13.5$ (see Figure 9 in \citealt{Evans18}), and because for $G < 12$, saturation starts to become an issue. The faint limit was chosen due to the flattening of the baseline curve, beyond which the scatter in magnitude uncertainty becomes too large for variable stars to be distinguished reliably.

Within each of these joint bivariate limits, we chose a random sample of 2000 stars, enabling a significant but tractable population of $\sim 1.5$ million stars overall to model the set of baseline curves. For example, we calculated one baseline curve for 2000 stars with $N_{obs,G} = 300-350$ and $16 < G < 16.1$, another for $N_{obs,G} = 350-400$ and $16 < G < 16.1$, etc., for the 770 bivariate bins. For each of these samples of 2000 stars, we calculated the mean magnitude uncertainty and the median absolute deviation, the latter as an estimate of the spread, robust to outliers.

Less than 0.2\% of stars in the \emph{Gaia} database have over 750 observations and their baseline curve is very similar to that for the stars with $N_{obs,G} = 700-750$, so that baseline curve is used for those stars with high numbers of observations.
However, for stars with fewer than 50 observations (corresponding to stars with typically 5 or fewer \emph{Gaia} focal plane transits), the scatter in the uncertainties was large and no well-defined baseline curve could be established. Therefore, stars with $N_{obs,G} <$ 50, which make up approximately 3\% of the \emph{Gaia} DR2 database, along with stars having $G < 14$ or $G > 19.5$, were excluded from the present work. Some high-amplitude variable stars might still be identifiable among the faint sources or those with few observations.

We also compared baselines at different Galactic latitudes. However, we found that Galactic latitude \emph{alone} has a negligible effect on the baseline curve. Galactic latitude indirectly affects our selection of variables because stars with close companions (in projection) can be shifted to higher photometric uncertainties even if not variable. The effects of crowding are addressed in Section \ref{sec:FalsePositives}.

\subsubsection{\label{sec:BP_RP_Phot} $G_{BP}$ and $G_{RP}$ Photometry}
1.4 billion sources in \emph{Gaia} DR2 that have reliable $G$ photometry also have independent blue ($G_{BP}$, hereafter $BP$) and red ($G_{RP}$, hereafter $RP$) photometry \citep{GaiaCollaboration18}.
Adding these measurements allows us to construct a subsample of our entire catalog of candidate variables that has lower completeness but higher purity.

\begin{figure}[t]
    \centering
    \includegraphics[width= \linewidth]{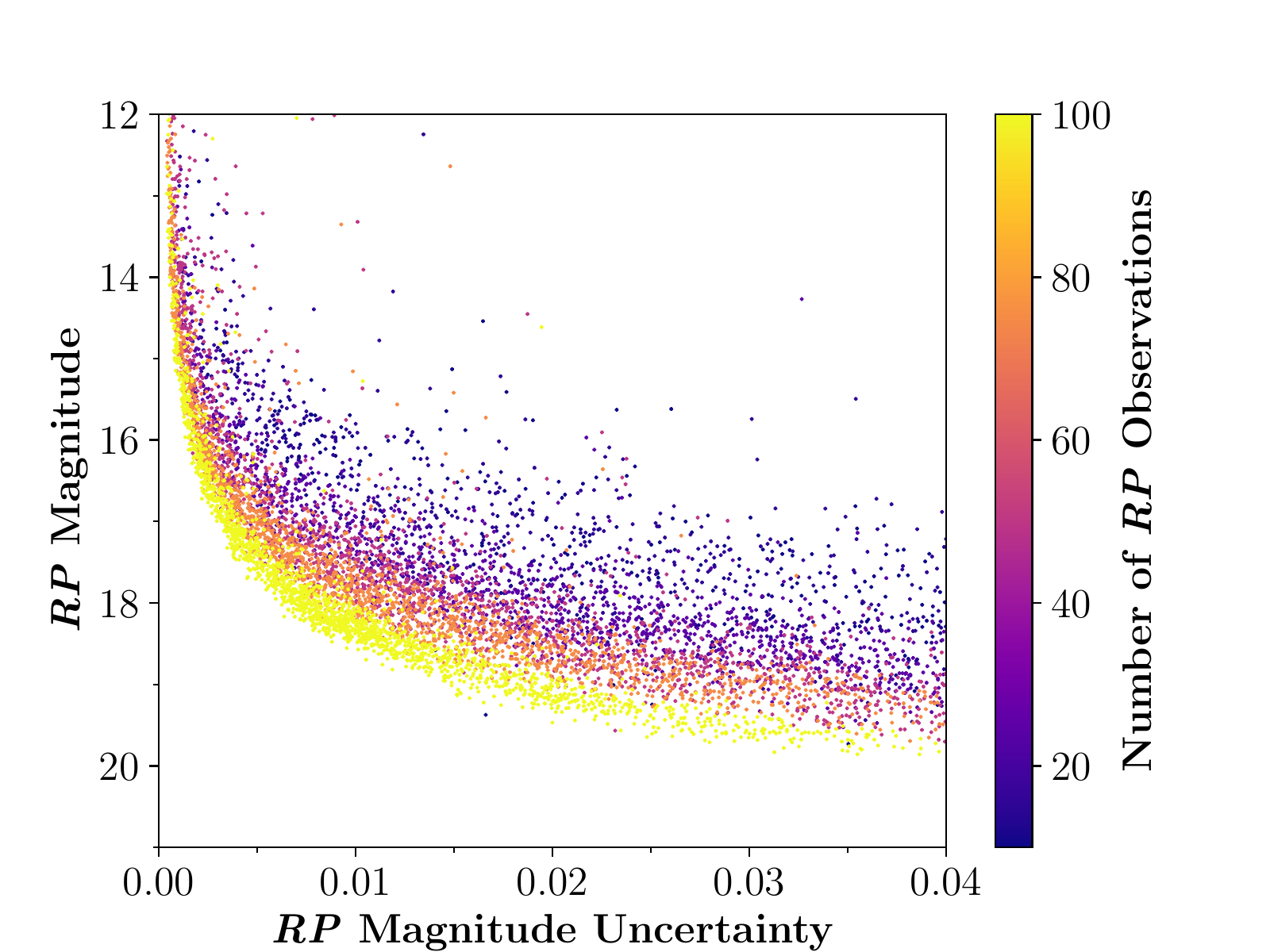}
    \caption{$RP$ mag vs. mag uncertainty plot, with the color representing the number of observations contributing to the \emph{Gaia} $RP$ photometry. The qualitative behavior is similar to that for $G$ (see Figure~\ref{fig:baselineCurves}).}
    \label{fig:RPBaseline}
\end{figure}

Figure \ref{fig:RPBaseline} illustrates that, as observed in $G$, $RP$ magnitude uncertainties decrease as the number of photometric observations increase (the figure for $BP$ is nearly identical and not shown). We constructed $BP$ and $RP$ baseline curves using a similar procedure as that for $G$ but tailored to the lower number of $BP$ and $RP$ observations. For each band, sources were binned by the number of observations, from 20 to 100 in intervals of 10, and binned again in magnitude from $BP$ (or $RP$) from 12 to 20 in intervals of 0.1 mag. For each of these 640 bivariate bins, we randomly selected 2000 stars to calculate the baseline curves from the mean magnitude uncertainty and median absolute deviation of each bin.
 
\subsection{\label{sec:FlaggingStars} Initial Candidate Variable Star Flagging}

We use the baseline curves to calculate an effective $\sigma$ for each bin in $G$ and $N_{obs,G}$, where 
\begin{equation}
    \sigma = 1.4826 \  \rm{ MAD} 
\end{equation}(the median absolute deviation). Then, for each source we can calculate the quantity $G\sigma$ as the ratio of the $G$ magnitude uncertainty in \emph{Gaia} DR2 for that source to the $\sigma$ for that bin.

\begin{figure}[t]
    \centering
    \includegraphics[width=\linewidth]{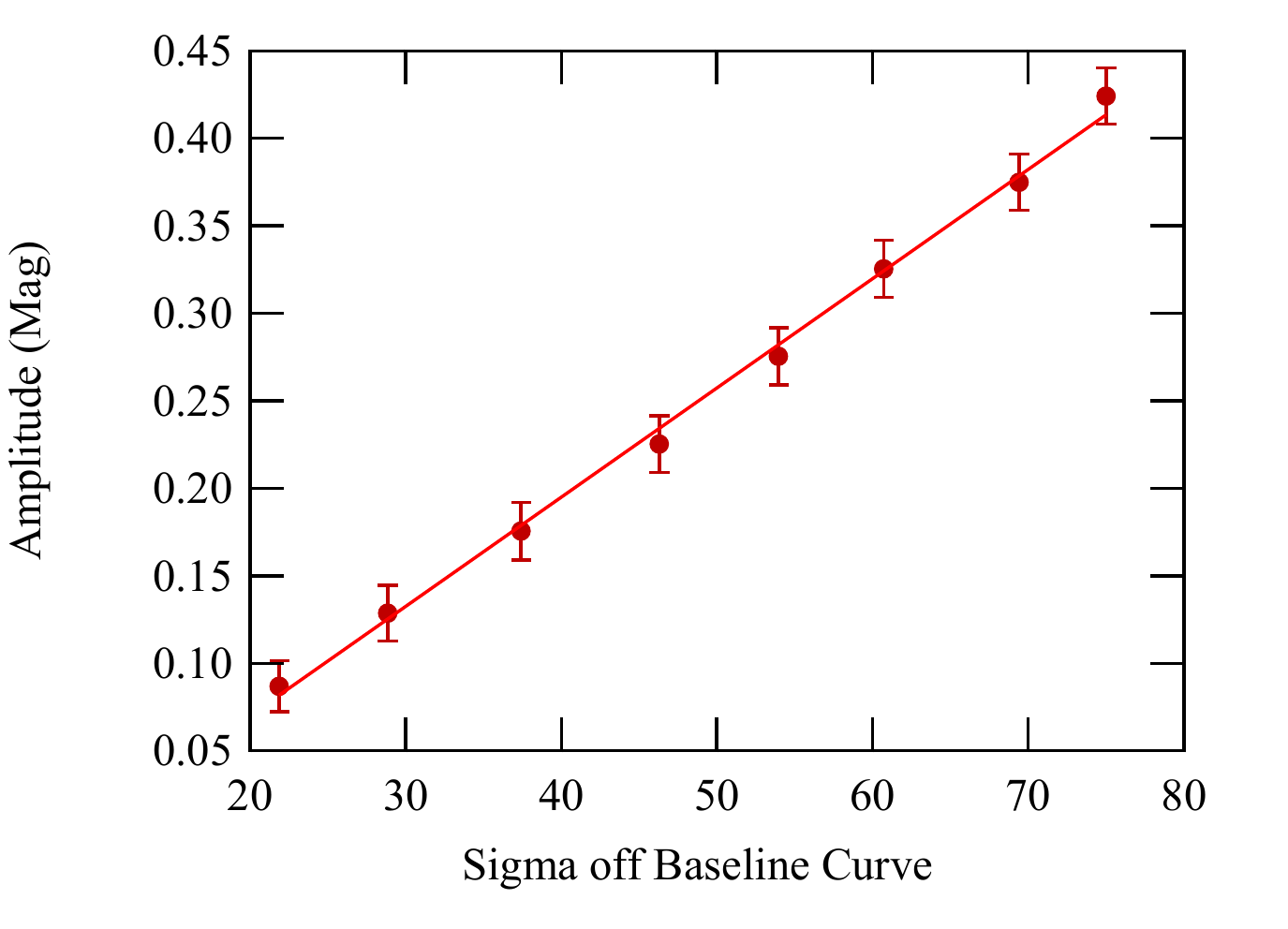}
    \caption{Amplitude of variability versus $G \sigma$ for CRTS stars with periods $< 10$ days. The plotted points are the mean amplitudes and $G \sigma$ of CRTS variables binned by amplitude in overlapping 0.1 mag bins, with bin centers from 0.1 to 0.45 mag. The uncertainties are given by standard errors of the mean, and the red line is a least-squares fit to these points                        to guide the eye.}
    \label{fig:MappingAmplitude}
\end{figure}

As evidence that $G\sigma$ tracks variability, Figure \ref{fig:MappingAmplitude} shows that the mean \emph{amplitude} of variability in the CRTS sample correlates with $G\sigma$: on average, stars that deviate more from the baseline curve have higher amplitude variability.

\begin{figure}[t]
    \centering
    \includegraphics[width=0.9\linewidth]{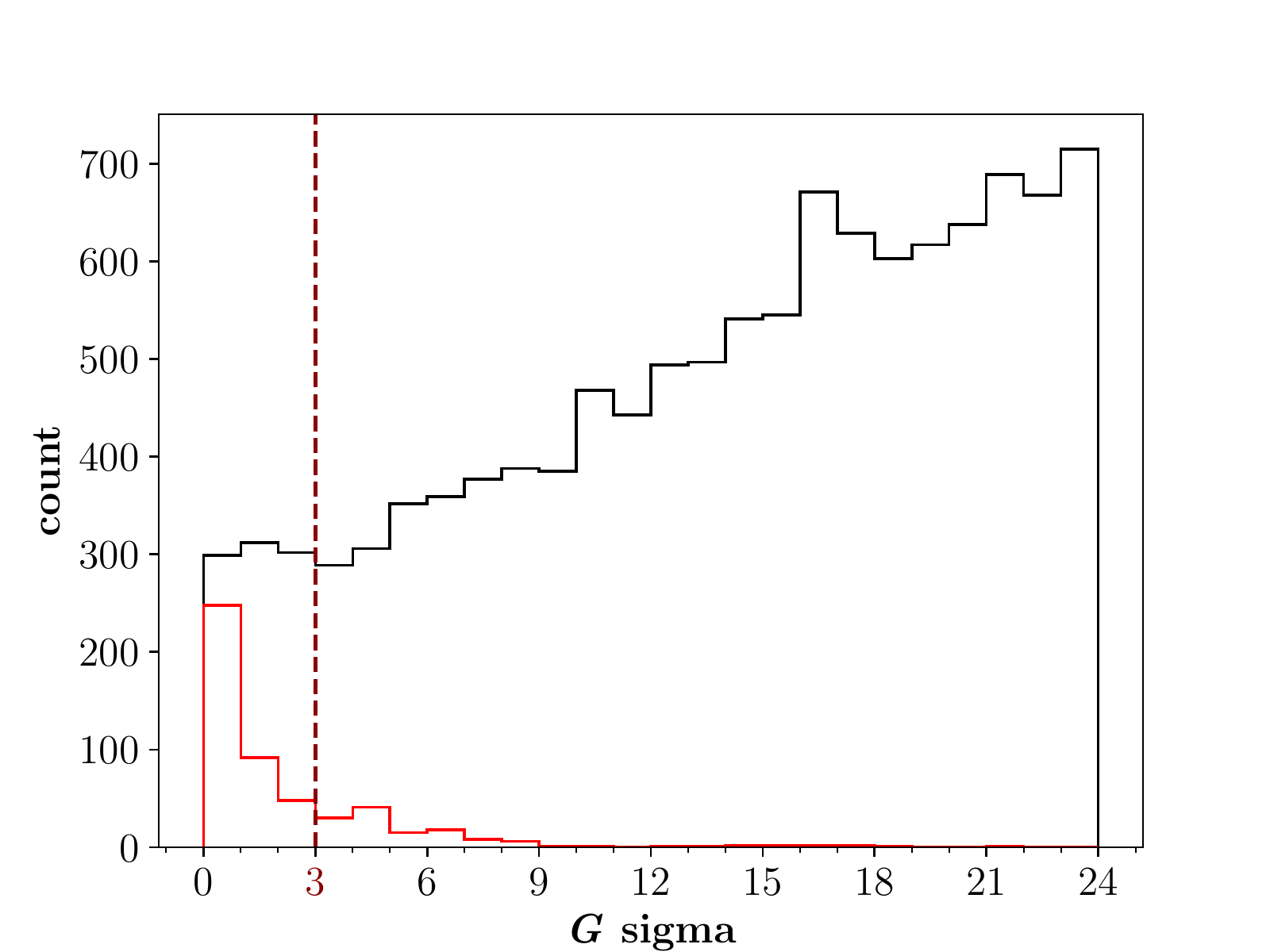}
    \caption{$G\sigma$ distribution of CRTS stars with $G\sigma < 24$ (black line). In red is the $G\sigma$ distribution for a random sample of 2000 stars.}
    \label{fig:GsigmaHist}
\end{figure}

We flagged an initial list of candidate variable stars as those stars with $G\sigma> 3$. This limit is justified by Figure \ref{fig:GsigmaHist}, which compares the lower tail of the $\sigma$ distribution for the CRTS variables (the sample discussed in \S 3.1) to a random sample of \emph{Gaia} sources. A $3\sigma$ cut includes 96\% of CRTS variables, while few randomly selected stars have $G\sigma > 3$. After running the algorithm across the entire sky, the initial list of  candidate variables with $G\sigma > 3$ contained \initialCatalogList~stars. Note that \fourGsig stars were flagged with $G\sigma > 4$, and \fiveGsig stars were flagged with $G\sigma > 5$. Because our catalog prioritizes identifying candidate variables where the existence of any such variable is useful, we have chosen a cut of $G\sigma > 3$: $G\sigma$ values are provided in our catalog if the user chooses to make stricter cuts.

\subsubsection{False Positives: Galaxies and Stars with Neighbors}
\label{sec:FalsePositives}

Two source classes can have $G\sigma > 3$ even if they are not variables. The first is resolved galaxies. We cross-checked our list of candidates with galaxies in the GLADE \citep{Dalya2018} and SDSS galaxy catalogs \citep{Simard_2011}, removing candidates that matched to sources in these catalogs to within $2\arcsec$.

The second class of potential ``false positives" are sources with higher-than-normal backgrounds, typically because they are close to other stars, especially bright ones. Careful assessment of the effects of neighboring stars led to the following criteria for flagging candidates as potential false positives: (i) Any neighboring star within $5\arcsec$; (ii) a neighboring star with $G < 20$ within $10\arcsec$; (iii) a neighboring star with $G < 12$ within $30\arcsec$; or (iv) a neighboring star with $G < 8$ within $50\arcsec$. A small subset of sources with close neighbors was included in the ``Bronze" sample discussed in the next subsection.

Of the initial \initialCatalogList~sources flagged for $G\sigma > 3$, \NeighborsRemoved~were removed as false positives. Of the remaining \CatalogNeighborsRemoved~sources, \GalaxiesRemoved~were removed after cross-matching with GLADE and SDSS galaxy catalogs.

\subsection{Adding $BP$ and $RP$: Gold, Silver, and Bronze Criteria}

For stars that have $BP$ and $RP$ magnitudes, we determine $BP\sigma$ and
$RP\sigma$ values from their respective baseline curves in the same manner as for $G\sigma$.

We define the Gold sample as 
those sources with all of $G\sigma >3$, $BP\sigma > 3$, and $RP\sigma > 3$ and that pass the nearby neighbor criteria. This sample will be the least complete but should have the highest proportion of true variable stars.

The Silver sample consists of those sources with $G\sigma >3$ and either $BP\sigma > 3$ or $RP\sigma > 3$ and which also pass the nearby neighbor criteria, but are not in the Gold sample.

The Bronze sample is the union of sources with (i) $G\sigma >3$ that pass the nearby neighbor criteria and do not fall into the Gold or Silver samples, and (ii) sources that do \emph{not} pass the nearby neighbor criteria, but nevertheless have all of $G\sigma >3$, $BP\sigma > 3$, and $RP\sigma > 3$.

We note that in the mean these three measurements are positively correlated with one another, as would be expected 
given their independent tracing of the variability. It is possible that a careful consideration of the relative $BP\sigma$ and $RP\sigma$ values for a given source could give interesting constraints on the nature of its variability, but in this paper the main utility of using multiple measurements is simply to average down the noise for fainter or lower amplitude variables.

\begin{figure}
    \centering
    \includegraphics[width=\linewidth]{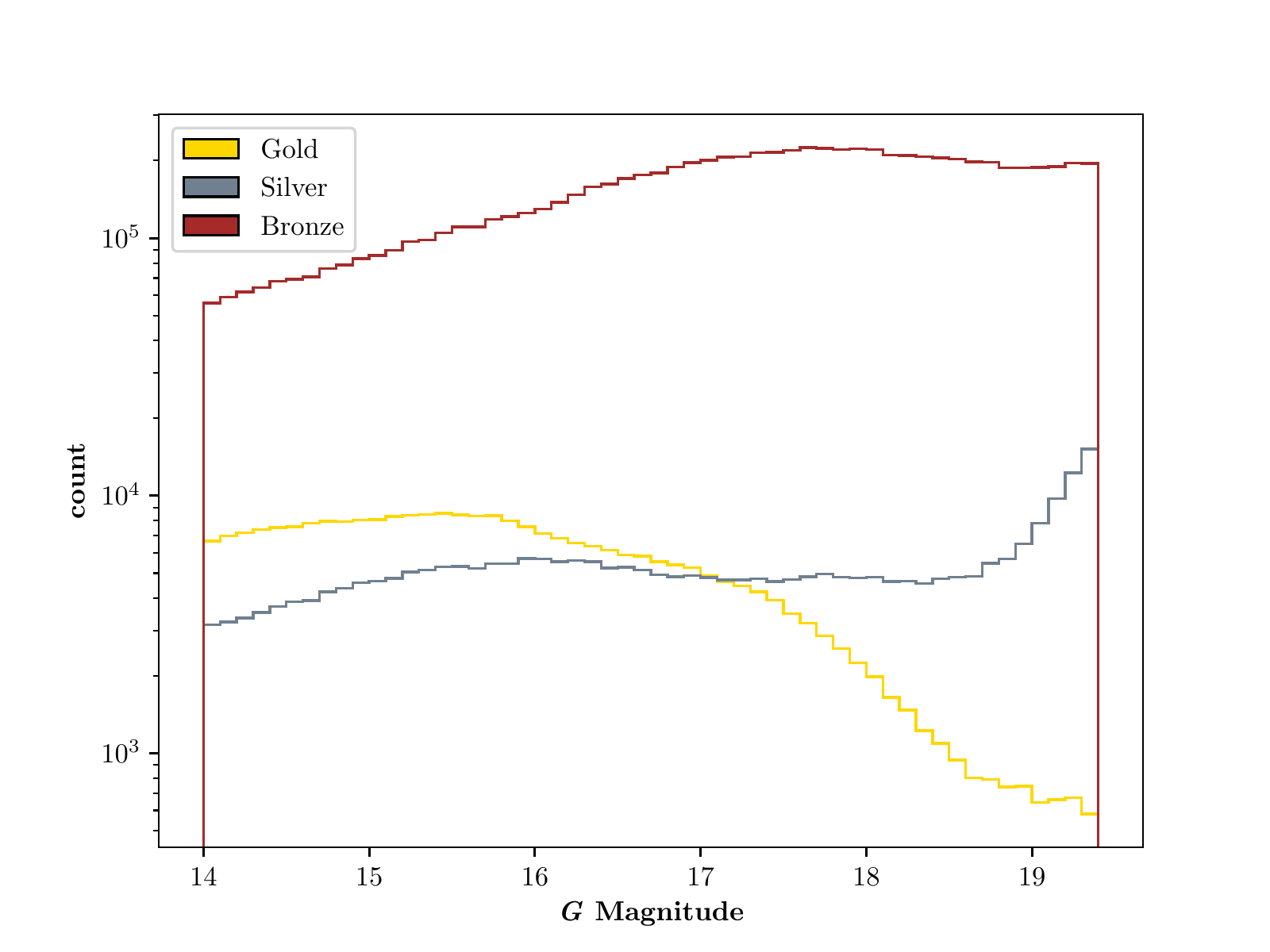}
    \caption{$G$ mag distribution of candidate variable stars, with color representing the Gold, Silver, or Bronze samples.}
    \label{fig:HistogramDistribution}
\end{figure}

The magnitude distribution of each of these samples is shown in Figure \ref{fig:HistogramDistribution}. The mean $G$ magnitudes of our Gold, Silver, and Bronze samples are 16.2, 17.4, 18.1, respectively. Note that our Gold sample consists of brighter sources because $RP$ and $BP$ photometry are available for fewer dim stars. Figure \ref{fig:HistogramDistribution} also shows an uptick in silver sources at around $G$ = 18.5, which is likely due to the larger scatter in $RP$ and $BP$ magnitude errors at fainter magnitudes as compared to $G$ magnitude errors (see Figures 3 and 4). We do not argue that the different samples represent different classes of variable objects, but rather that
the Gold, Silver, and Bronze samples primarily represent different levels of confidence that real variability is present.

\subsection{Our Full Catalogs}
\label{sec:fullcatalog}
Our full catalog consists of \FinalTotalStarsFlagged~candidate variables with $14 < G < 19.5$.
As shown in Table \ref{table:summary}, the majority of these (\FinalBronzeFlagged) fall into the Bronze sample, with \FinalSilverFlagged~and \FinalGoldFlagged~sources in the Silver and Gold samples, respectively. A sample of 50 candidates is listed in 
Table \ref{tab:CandidateVariablesTable}, and we provide the full catalog in machine-readable format.

The comparisons to existing catalogs in Sec. \ref{sec:Catalina} and \ref{sec:Strip82} show that our method can identify periodic variables with shorter periods and higher amplitudes with fidelity, with some sensitivity to longer period or irregular variables. Figure \ref{fig:16CandidateVariables} presents the CRTS light curves of 15 new candidate variables contained in our catalog that are not present in any of the aforementioned variable star catalogs or any variable star catalogs in the Vizier\footnote{http://vizier.u-strasbg.fr/viz-bin/VizieR} database. As another illustration of science enabled by our catalog, we used this method to find the variable optical counterpart to the new likely redback millisecond pulsar 4FGL J2333.1--5527 \citep{Swihart2019ANL}, and this source is in our final catalog.

We emphasize that for non-periodic variables (such as young stellar objects, flare stars, and some background active galactic nuclei) our catalog is certainly incomplete and that the recovery fraction for these sources is very uncertain. Additionally, we note that the outlier rejection procedure used in the photometric processing for DR2 was sub-optimal for variables, with un-rejected faint outliers resulting in significantly dimmer mean magnitudes for a small fraction of sample variables (see \citealt{Arenou_2018}). Because a dimmer mean magnitude will also artificially raise the $3 G\sigma$ flagging threshold, we anticipate some variables with un-rejected faint outliers will be missed in our catalog.

\begin{deluxetable}{lr}
\tablewidth{\linewidth}
\tablecaption{A Summary of our Catalog of\\
Candidate Variable Stars in \emph{Gaia} DR2}
\tablehead{\large{Catalog Subset} & \large{\# of Sources}}
\startdata
Full Catalog & \FinalTotalStarsFlagged\\[3pt]
\hline\\[-3pt]
Bronze\tablenotemark{a} & \FinalBronzeFlagged\\[3pt]
Silver\tablenotemark{b} & \FinalSilverFlagged\\[3pt]
Gold\tablenotemark{c} & \FinalGoldFlagged\\[3pt]
\hline\\[-3pt]
$G\sigma > 3$\tablenotemark{d} & \initialCatalogList\\[3pt]
Stars removed due to neighbor criteria\tablenotemark{e} & \NeighborsRemoved \\
Stars removed after galaxy cross-matching \tablenotemark{f} & \GalaxiesRemoved
\enddata
\tablenotetext{a}{Only $G\sigma > 3$ with no nearby neighbor or $G\sigma,RP\sigma,BP\sigma > 3$ with a nearby neighbor. }
\tablenotetext{b}{$G\sigma,RP\sigma > 3$ or $G\sigma,BP\sigma > 3$, with no nearby neighbor }
\tablenotetext{c}{$G\sigma,RP\sigma, BP\sigma > 3$, with no nearby neighbor}
\tablenotetext{d}{See Sec.~\ref{sec:BP_RP_Phot}}
\tablenotetext{e}{See Sec.~\ref{sec:FalsePositives}}
\tablenotetext{f}{Number of sources matched to GLADE and SDSS galaxy catalogs within 2". See Sec.~\ref{sec:FalsePositives}}
\label{table:summary}
\end{deluxetable}

\section{Results}
\label{sec:Results}

We tested our method for selecting candidate variable stars and our subsamples in several ways.

\subsection{CRTS Periodic Variables}
\label{sec:Catalina} 

Our first test is for periodic variables of relatively short period, using a combination of the northern and southern CRTS Periodic Variables Catalogs \citep{Drake14,Drake17}. Together these catalogs contain 70,680 periodic variables with periods $< 10$ d and magnitudes $14 < G < 19.5$. We find that our Gold selection recovers \CatalinaGoldFlagged~of the CRTS variables, with the Silver and Bronze samples adding \CatalinaSilverFlagged~and \CatalinaBronzeFlagged~sources, respectively. The total Bronze+Silver+Gold sample recovers \CatalinaGsigmaFalsePositivesWithBPandRP~(92\%) of the CRTS variables.  We note that \CatalinaTotalFlagged~(96\%) of the variables had $G\sigma>3$; hence the nearby neighbor cut led to only a 4\% loss in the recovery of variable stars. This loss estimate is likely optimistic for our full-sky catalog since CRTS avoids the dense Galactic Plane.

\begin{figure}[t]
    \centering
    \includegraphics[width=\linewidth]{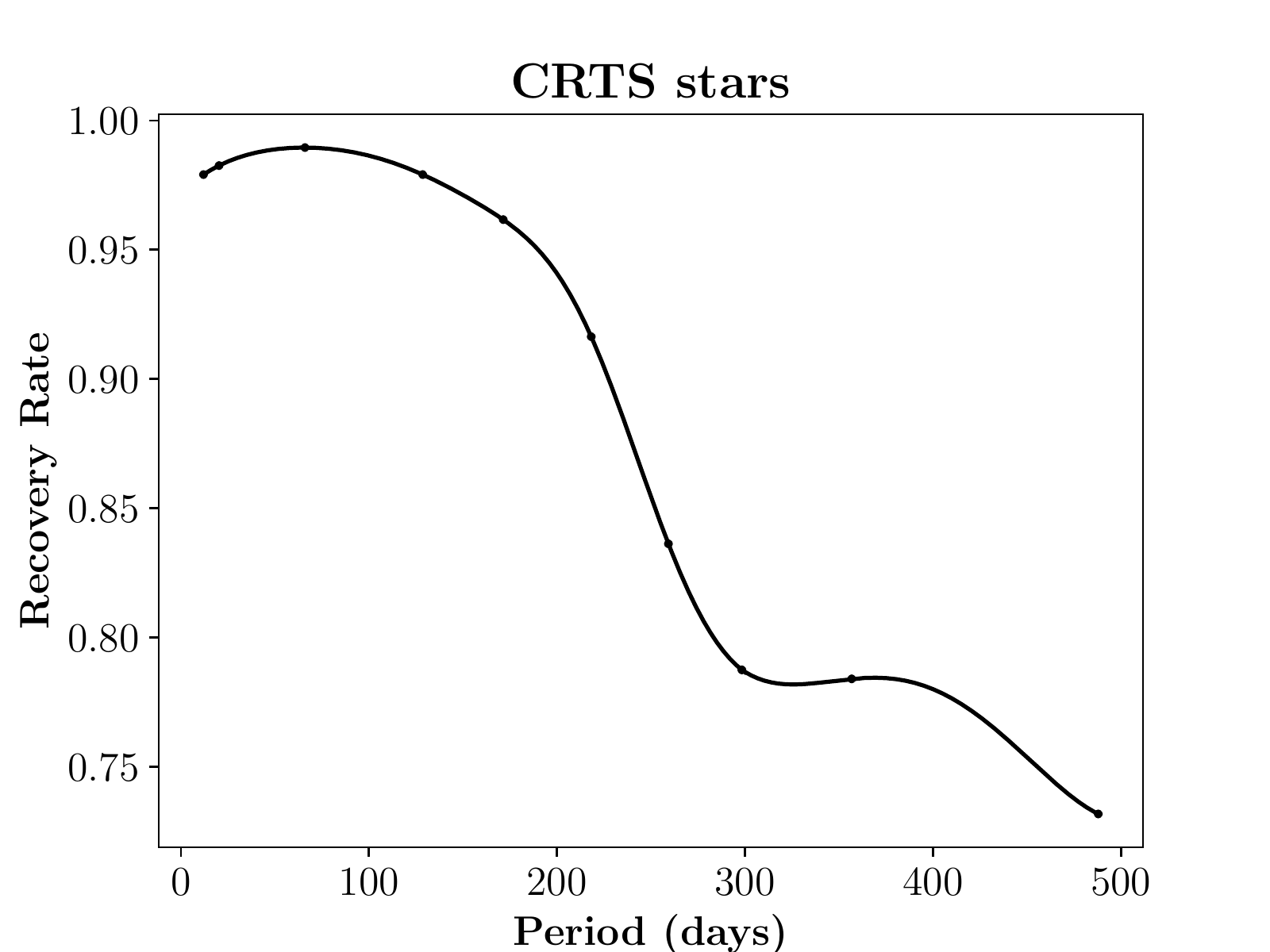}
    \caption{Recovery rate versus photometric period for binned CRTS stars with periods $>$ 10 days and magnitudes $14 < G < 19.5$. 2,870 CRTS stars were sorted by period and binned so that an equal number of stars were in each bin. Plotted is the mean recovery rate and period of each bin; note that the recovery rate decreases with increasing period.}
    \label{fig:CatalinaPeriods}
\end{figure}

Given the still limited time range and number of visits in DR2 compared to the expected final values, we expect that our variable selection method will be less effective for long-period variables. To test this, we expand our analysis of the CRTS variable star sample to longer periods. Figure \ref{fig:CatalinaPeriods} shows the  recovery rate of CRTS variables as a function of period. While still above 95\% out to periods of half a year, it decreases gradually for periods $\gtrsim 100$ d, reaching a recovery rate of $\lesssim 75\%$ for periods longer than 1 yr. We expect that future data releases will allow improved recovery of periodic variables with longer periods using a similar methodology.

\subsection{Gaia Periodic Variables}
\label{sec:GDR2_RRlyraeCepheids} 

As another test of our catalog, we also compared it to the initial catalog of RR Lyrae and Cepheid variables provided in \emph{Gaia} DR2 \citep{GaiaCollaboration18}. The Cepheid variable catalog contained \CepheidTotal sources with magnitudes $14 < G < 19.5$. Our Gold selection recovers \CepheidGoldFlagged of the Cepheid variables, while the Silver and Bronze selections recover \CepheidSilverFlagged and \CepheidBronzeFlagged, respectively. These selections in total recover \CepheidFlagged (91\%) of the \CepheidTotal variables. However, we note that 100\% of the variables had $G\sigma>3$.

The RR Lyrae variable catalog contained \RRLyraeTotal~sources with magnitudes $14 < G < 19.5$. Our Gold selection recovers \RRLyraeGoldFlagged~of the RR Lyrae variables, while the Silver and Bronze selections recover \RRLyraeSilverFlagged~and \RRLyraeBronzeFlagged. These selections in total recover \RRLyraeFlagged~(71\%) of the \RRLyraeTotal~variables. \RRLyraeGFlagged~(99.8\%) of the variables had $G\sigma>3$.

These comparisons indicate that the photometric uncertainty selection ($G\sigma>3$) correctly picks out nearly all of the \emph{Gaia} DR2 Cepheids and RR Lyrae variables so far identified, and that the main source of incompleteness is the nearby neighbor cuts needed to exclude false positives.

\subsection{Zwicky Periodic Variables}
\label{sec:Zwicky} 

The \emph{Zwicky} variable catalog contains \ZwickyTotal~sources with magnitudes $14 < G < 19.5$ and periods $<$ 10 d. Our Gold selection recovers \ZwickyGoldFlagged~of the Zwicky variables, while the Silver and Bronze selections recover \ZwickySilverFlagged~and \ZwickyBronzeFlagged, respectively. These selections in total recover \ZwickyFlagged (65\%) of the \ZwickyTotal~variables.  However, we note that \ZwickyGFlagged~(94\%) of the variables had $G\sigma>3$, so again a substantial number of true variables are excluded from our catalog due to having nearby neighbors. 

\subsection{Stripe 82}
\label{sec:Strip82}

Stripe 82 is a region of sky with many observations from SDSS and hence one in which variability has been relatively well-characterized.

\subsubsection{Existing Catalogs}

There are several existing catalogs of variables in the Stripe 82 region, including both Galactic variable stars and other variable sources, such as background active galactic nuclei.

\citet{Sesar_2007} identified 13,051 candidate variable sources in Stripe 82 with root mean square variability of at least 0.05 mag in $g$ and $r$. Of the 5476 sources in our magnitude range that pass our false positive selection criteria (not identified as a galaxy or with nearby neighbors), 564 (10\%), 236 (4\%), and 1713 (31\%) are recovered in the Gold, Silver, and Bronze samples respectively, for an overall recovery fraction of 46\%.

In the enlarged SDSS Stripe 82 Variable Source Catalog, constructed in a similar manner to \citet{Sesar_2007} but with observations from both SDSS-I and SDSS-II \citep{Ivezi07}, there are 24,719 sources that pass our criteria for evaluation as variables (see Section 3.4), with respective Gold, Silver, and Bronze sample recoveries of 688 (3\%), 308 (1\%), and 2454 (10\%). The summed recovery percentage is only 14\%.

Contrasting with the low recovery rate from this full catalog of potential variables, we also looked at the \citet{Sesar2010} sample of RR Lyraes selected from \citet{Ivezi07}. Of these 483 RR Lyraes, 419 meet our criteria, and of these, we recover 59\% in our Gold sample and 406/419 (97\%) in the combined Gold+Silver+Bronze selection.

The ATLAS survey has also cataloged candidate variable stars down to a magnitude limit generally consistent with our \emph{Gaia} limits over a wide area that includes Stripe 82 \citep{Heinze_2018}. We first consider their sample of ``probable" variables, which has 800 sources in Stripe 82 that pass our magnitude and other cuts; we recover 253 (32\%), 352 (44\%), and 114 (14\%) in the respective Gold, Silver, and Bronze samples, with a combined recovery rate of 90\%. If we expand to their larger sample of 5725 ``dubious" variables, which includes sources whose variability is less certain, we recover 827 (14\%), 323 (6\%), and 679 (12\%), for a combined recovery rate of 32\%.

Both the CRTS results (Sec. \ref{sec:Catalina}) and those for Stripe 82 are consistent with a high recovery fraction for probable variable stars, especially those with relatively short periods and/or higher amplitudes. The status of the sources not identified as variable using our \emph{Gaia} method is less clear. 
Some of the Stripe 82 objects have CRTS light curves, and an examination of a random sample of these light curves shows little or no evidence for variability in many cases. It may be that these are periodic variable stars with periods too long to show substantial variability in \emph{Gaia} DR2, irregular variable stars, or distant background sources such as active galactic nuclei with no variability in DR2.
This does not account for all the variables found in previous surveys and not recovered by our method. For example, some of the missing CRTS sources have been classified as periodic variables. We closely examined light curves for hundreds of these objects and confirmed that they do indeed appear to be variable, but found no trend in variable type, period, or amplitude suggesting why these sources were not flagged as having $G\sigma > 3$. In some cases it may be due to a small total number of $G$ observations and that these happen to fall during orbital phases with less or no photometric variation. It will be worthwhile to investigate these sources using future \emph{Gaia} data releases with a longer time baseline and a larger number of observations.

\subsubsection{New Variable Candidates in Stripe 82}

Now we discuss the results from our method. 
In Stripe 82, our Bronze sample contains \StripeBronzeNEWflagged~sources: \StripeBronzeGalaxiesRemoved~with $G\sigma > 3$ and no nearby neighbors and 798 with nearby neighbors but all of $G\sigma > 3$, $BP\sigma > 3$, and  $RP\sigma > 3$. The Silver and Gold samples are much smaller: \StripeSilverGalaxiesRemoved~and
\StripeGoldGalaxiesRemoved, respectively.

\begin{figure}[t]
    \centering
    \includegraphics[width=0.9\linewidth]{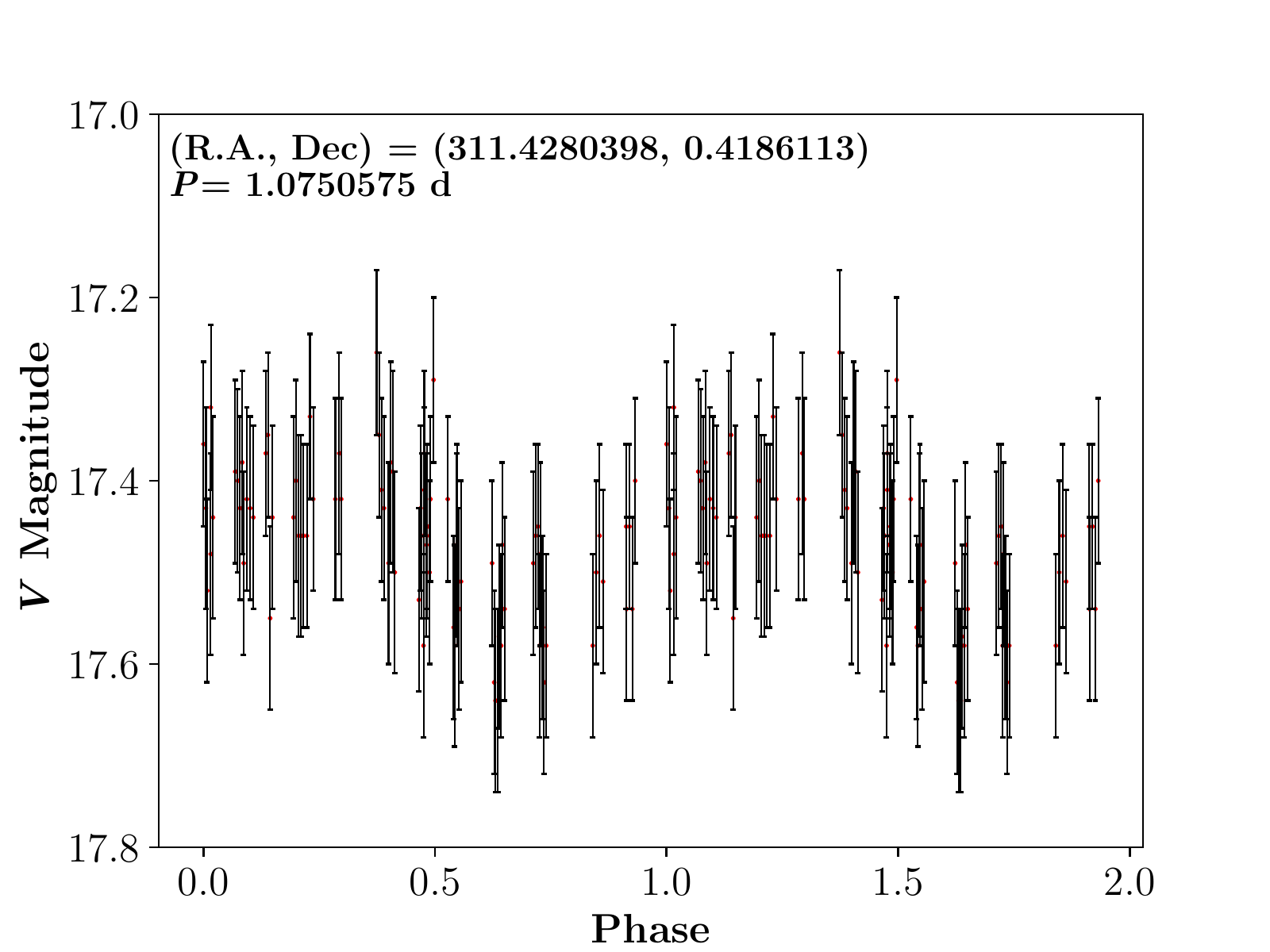}
    \caption{Light curve of a new candidate variable star in Stripe 82 located at (R.A., Dec) = (311.4280398, 0.4186113) identified by our algorithm, but not present in the ATLAS or SDSS Stripe 82 Variable Source Catalogs. Light curve obtained from CRTS \citep{Drake2009}.} 
    \label{fig:StripeSampleVariable}
\end{figure}

\begin{figure*}
\centering
\includegraphics[width=2.2\columnwidth]{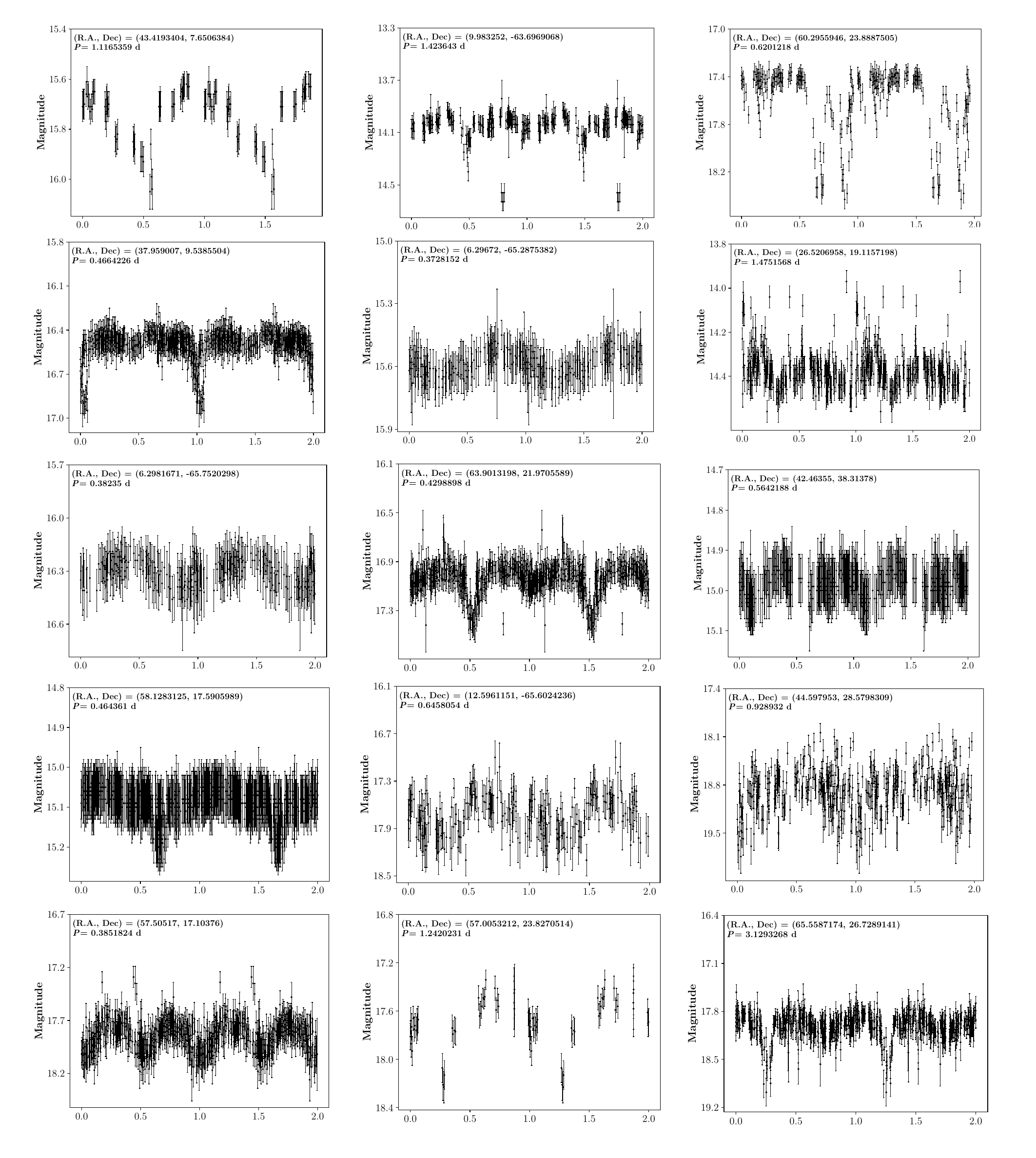}
\caption{Example light curves of 15 new candidate variables, not present in any known variable star catalogs. Light curves obtained from CRTS \citep{Drake2009} and folded on the candidate periods listed.}
\label{fig:16CandidateVariables}
\end{figure*}

Of the \StripeGoldGalaxiesRemoved~sources in the Gold sample, 743 are also found in the SDSS Stripe 82 Variable Source Catalog or in the ATLAS variable catalog \citep{Ivezi07,Heinze_2018}, an overlap of 91\%. The overlap fraction is smaller for the Silver sample (50\%) and yet smaller for the Bronze sample, where only 3624 (20\%) of the \StripeBronzeNEWflagged~sources are in the SDSS Stripe 82 or ATLAS catalogs. Of the 77 variables in the Gold Catalog not present in the SDSS Stripe 82 or ATLAS variable catalogs, CRTS light curves were available for 74. An inspection of CRTS light curves for the 74 new variable star candidates shows that some are indeed true variables (see example in Figure \ref{fig:StripeSampleVariable}). Based on the light curves, we estimate that 51 of the variable candidates are false positives, making up $\sim$6\% of the \StripeGoldGalaxiesRemoved~sources in the Gold sample. 

\subsection{Test of a Random Sample}

As an additional test, we randomly chose 1000 sources from our catalog of candidate variables in approximate proportions to the relative sample sizes (30 Gold, 60 Silver, and 910 Bronze), with the only proviso that they have CRTS data to allow us to examine their light curves. Of these 1000 sources, 846 (85\%) show clear variability in the CRTS data. This comparison suggests that up to $\sim 15\%$ of our catalog might be false positives, though it is also possible that some of these sources are true non-periodic variables that happen to not show variability over the time range of the CRTS light curves.

\subsection{Preliminary Comparison to \citet{Mowlavi2020}}

After submission of the present paper, a preprint was posted to the arXiv \citep{Mowlavi2020} that employs a similar method to this paper in identifying candidate variables, finding a catalog of 23.3 million candidates (compared to our 9.3 million). A full comparison is outside of the scope of this paper, but we give a brief comparison here. We find that 2.8 million of our 9.3 million candidates are listed as variables in that catalog. Of our 6.5 million candidate variables not listed in \citet{Mowlavi2020}, we find that 368,717 cross-match with some of the existing variable star catalogs discussed above, including those from CRTS, ATLAS, and Zwicky, which suggests that there are areas of parameter space where our catalog is more complete. Further, we find that of their full catalog of 23.3 million sources, 
19.5 million were in our preliminary catalog 
 \emph{before} we removed nearby neighbors, suggesting a large part of the difference between the catalogs concerns the treatment of sources with nearby neighbors. \citet{Mowlavi2020} also includes analysis not present in this paper, including using the relative evidence for variability in the $RP$ and $BP$ bands to help classify the candidate variables.

\section{\label{sec:Conclusion} Conclusions and Future Work}
We have presented a catalog of over 9 million candidate variable stars identified by their large photometric uncertainties. Our catalog is divided into three samples (Gold, Silver, and Bronze) based on probability of variability. After cross-matching with several published variable catalogs, we estimate our recovery rate to be high ($>90\%$) for stars in non-crowded regions with periods $<10d$. After cross-matching in Stripe 82 and subsequently investigating light curves, we also estimate the purity of our Gold sample to be $94\%$. 

Epoch photometry for all sources observed from \emph{Gaia} will not be available until 2021 at the earliest. Until then, indirect methods of probing variability across the whole sky, such as the algorithm presented in this paper, can be useful. The longer time baseline of \emph{Gaia}'s Early Data Release 3, expected in late 2020, should allow the improved detection of variables though application of a similar analysis.

\section*{Acknowledgements}

We thank the anonymous referee whose detailed comments improved the quality, clarity, and precision of this work.

We acknowledge support from the Packard Foundation and NSF grant AST-1714825.

This work has made use of data from the European Space Agency (ESA) mission {\it Gaia} (\url{https://www.cosmos.esa.int/gaia}), processed by the {\it Gaia} Data Processing and Analysis Consortium (DPAC, \url{https://www.cosmos.esa.int/web/gaia/dpac/consortium}). Funding for the DPAC has been provided by national institutions, in particular the institutions participating in the {\it Gaia} Multilateral Agreement. 

\bibliography{main}

\clearpage
\begin{turnpage}
\begin{deluxetable}{lllccccccccccccc}
\tablecaption{Parameters of Candidate Variables}
\tablehead{DR2 source ID & RA\_ICRS & DEC\_ICRS & $G_{\rm{mag}}$ & $G_{\rm{err}}$ & $N_{obs,G}$ & $G\sigma$ & $RP_{\rm{mag}}$ & $RP_{\rm{err}}$ & $N_{obs,RP}$ & $RP\sigma$ &$BP_{\rm{mag}}$ & $BP_{\rm{err}}$ & $N_{obs,BP}$ & $BP\sigma$ & Neighbor}
\startdata
40439865874173824 & 59.6601591 & 16.3837984 & 19.1279& 0.01492 & 221 & 26.50&17.7657&0.01914 & 22&0.20 & 19.8941&0.01914&21 & 0.33 & 0 \\ 
44826779826993280 & 55.2163228 & 18.0217655 & 18.0111& 0.00577 & 255 & 25.98&17.1618&0.01729 & 28&1.92 & 18.6785&0.01729&27 & 0.01 & 0 \\ 
46738044570131712 & 61.9564343 & 17.2009303 & 19.4670& 0.01528 & 158 & 15.38&17.9242&0.06463 & 14&2.37 & 20.4754&0.06463&13 & 2.87 & 0 \\ 
48486989612456832 & 63.0898901 & 18.6801123 & 16.3860& 0.00386 & 213 & 28.20&15.2971&0.00354 & 21&0.72 & 17.3459&0.00354&21 & 0.06 & 0 \\ 
58445060267493632 & 53.8329090 & 21.2487259 & 16.6516& 0.00295 & 140 & 7.08&15.4208&0.03208 & 15&11.04 & 16.8823&0.03208&16 & 21.56 & 1 \\ 
86177389218730496 & 38.5010657 & 18.8148091 & 14.9954& 0.00223 & 124 & 10.72&14.3205&0.00564 & 14&1.81 & 15.5253&0.00564&14 & 1.86 & 0 \\ 
117100466555852800 & 50.6010598 & 24.8243226 & 17.1271& 0.00164 & 235 & 3.48&15.7194&0.01731 & 24&13.60 & 17.9888&0.01731&23 & 8.21 & 1 \\ 
143145251318785920 & 44.3635758 & 38.3653871 & 17.1593& 0.02211 & 143 & 85.55&15.8443&0.00907 & 16&0.19 & 17.6432&0.00907&16 & 0.25 & 0 \\ 
148930709705115904 & 64.0776437 & 23.2004018 & 16.1915& 0.00490 & 260 & 51.33&14.8645&0.00555 & 33&7.82 & 16.5142&0.00555&36 & 4.97 & 1 \\ 
151237485099759232 & 66.5508633 & 25.9898104 & 14.4475& 0.00074 & 211 & 3.42&13.4180&0.00276 & 22&2.56 & 15.5048&0.00276&23 & 2.77 & 0 \\ 
153617442441325184 & 72.8391093 & 26.3817344 & 18.4952& 0.00341 & 292 & 5.98&17.4057&0.01234 & 30&0.03 & 19.3327&0.01234&31 & 1.04 & 0 \\ 
154831032697082368 & 73.0370062 & 27.5345148 & 16.4009& 0.00110 & 296 & 3.37&15.3566&0.00418 & 31&2.01 & 17.2968&0.00418&31 & 0.68 & 0 \\ 
158823496855936384 & 68.5318062 & 29.0404116 & 18.3049& 0.00485 & 193 & 6.21&17.0001&0.02587 & 21&5.33 & 18.3182&0.02587&22 & 3.20 & 1 \\ 
160235647742062976 & 71.9329177 & 31.2417966 & 19.3166& 0.00487 & 285 & 3.00&18.1348&0.02604 & 29&0.12 & 20.4003&0.02604&27 & 2.78 & 0 \\ 
161013616643339520 & 71.8608366 & 31.6261585 & 18.9280& 0.00394 & 287 & 3.60&17.9972&0.02629 & 25&0.86 & 19.5425&0.02629&28 & 1.25 & 0 \\ 
162749367547364352 & 63.5556061 & 26.6509968 & 17.1580& 0.00544 & 219 & 34.67&15.8288&0.00655 & 20&1.96 & 18.0270&0.00655&21 & 1.04 & 0 \\ 
167529636084675968 & 60.2216751 & 30.0819135 & 15.0750& 0.00077 & 443 & 7.70&14.0424&0.00475 & 39&10.03 & 15.9953&0.00475&42 & 6.02 & 1 \\ 
174062865455998464 & 71.5737286 & 35.1224336 & 16.4165& 0.00135 & 324 & 6.96&15.4715&0.00371 & 33&0.82 & 17.2830&0.00371&33 & 0.55 & 0 \\ 
182092182262153216 & 79.7161673 & 34.6546606 & 19.3924& 0.00860 & 113 & 3.13&18.1737&0.04528 & 12&0.39 & 19.8219&0.04528&12 & 0.02 & 0 \\ 
191485756774370432 & 87.0439433 & 39.4491555 & 19.1127& 0.00964 & 141 & 5.40&17.7361&0.03120 & 15&0.14 & 20.7538&0.03120&16 & 1.50 & 0 \\ 
200480655246963328 & 75.5775918 & 39.0991264 & 18.7465& 0.00399 & 210 & 3.62&17.8090&0.01975 & 22&0.02 & 19.7382&0.01975&21 & 0.70 & 0 \\ 
203855610478434176 & 69.1737776 & 41.2508518 & 19.1515& 0.00726 & 207 & 8.48&17.7001&0.05568 & 17&2.20 & 18.6667&0.05568&16 & 1.20 & 0 \\ 
219852783110541056 & 56.9022158 & 35.3482434 & 14.1686& 0.00055 & 413 & 6.35&13.6279&0.00202 & 45&2.90 & 14.5319&0.00202&46 & 2.05 & 0 \\ 
227679106878611328 & 65.4274274 & 40.3566799 & 18.3717& 0.00309 & 277 & 5.77&17.5991&0.01377 & 30&0.17 & 19.0430&0.01377&34 & 1.64 & 0 \\ 
233789608386812416 & 64.0863131 & 46.2473880 & 16.7640& 0.01450 & 213 & 123.65&15.4696&0.00518 & 23&2.37 & 17.2599&0.00518&24 & 1.35 & 0 \\ 
266688228946846848 & 79.9534749 & 54.8495438 & 18.1489& 0.00627 & 220 & 19.27&17.4057&0.01794 & 22&1.01 & 18.7789&0.01794&23 & 0.35 & 0 \\ 
335563050354523264 & 42.7281710 & 39.3564237 & 17.0391& 0.00604 & 157 & 27.75&16.3655&0.01924 & 16&1.86 & 17.5807&0.01924&16 & 1.91 & 0 \\ 
357698521524964096 & 28.8992244 & 49.7254524 & 15.1485& 0.00108 & 209 & 7.00&14.4436&0.00319 & 22&1.78 & 15.6625&0.00319&22 & 0.50 & 0 \\ 
378601062200955136 & 5.1365342 & 36.9321169 & 18.7991& 0.00410 & 342 & 8.06&17.7208&0.01464 & 40&0.66 & 19.7990&0.01464&39 & 0.50 & 0 \\ 
390288011812738560 & 12.0939593 & 47.7519071 & 17.2524& 0.00732 & 196 & 31.64&16.5988&0.00967 & 20&1.26 & 17.3824&0.00967&20 & 0.35 & 0 \\ 
403774346560833152 & 19.1895076 & 51.1509764 & 19.2292& 0.00953 & 236 & 13.53&17.2450&0.01446 & 21&0.57 & 18.7825&0.01446&23 & 0.34 & 0 \\ 
409865056863583616 & 21.0002057 & 51.7827563 & 15.6915& 0.00464 & 271 & 50.42&14.8769&0.00301 & 32&1.77 & 15.9502&0.00301&32 & 0.12 & 0 \\ 
410463947104223104 & 20.9543133 & 53.3759829 & 17.3462& 0.00156 & 270 & 3.36&16.6246&0.00912 & 23&0.54 & 17.8409&0.00912&24 & 0.05 & 0 \\ 
412647469820585856 & 23.3232320 & 56.0670938 & 18.6842& 0.00344 & 483 & 9.63&17.5507&0.01151 & 46&0.06 & 19.3642&0.01151&46 & 0.22 & 0 \\ 
418061469380734976 & 8.3388897 & 54.0476103 & 17.9764& 0.00159 & 765 & 7.80&17.1787&0.00615 & 76&0.27 & 18.4924&0.00615&79 & 0.90 & 0 \\ 
419008144586789504 & 6.3448846 & 52.6580333 & 19.3008& 0.00846 & 391 & 18.04&18.2719&0.02831 & 37&0.94 & 19.7138&0.02831&38 & 1.50 & 0 \\ 
421811830519596160 & 7.4454725 & 57.0624586 & 19.1549& 0.00570 & 318 & 9.78&17.9427&0.02408 & 30&1.25 & 19.7620&0.02408&29 & 2.58 & 0 \\ 
425675724840727424 & 15.1997316 & 58.2080534 & 16.5802& 0.00151 & 408 & 14.03&15.7121&0.00496 & 47&2.93 & 17.3134&0.00496&44 & 2.55 & 0 \\ 
428039155142976768 & 7.7996961 & 58.9323984 & 19.2804& 0.00603 & 357 & 10.66&18.3513&0.02880 & 34&0.68 & 20.0932&0.02880&36 & 0.48 & 0 \\ 
431237329533972480 & 4.1228458 & 62.1501714 & 17.6958& 0.00290 & 286 & 11.96&16.8993&0.01099 & 27&0.84 & 18.3370&0.01099&27 & 1.32 & 0 \\ 
431740527902685568 & 0.4187533 & 64.3536029 & 17.2727& 0.00193 & 297 & 6.95&16.3116&0.00682 & 29&0.26 & 18.1417&0.00682&28 & 0.96 & 0 \\ 
436018688979761280 & 45.9688770 & 47.7504148 & 15.6459& 0.00249 & 199 & 9.32&14.9966&0.00331 & 20&1.09 & 16.0294&0.00331&20 & 1.15 & 0 \\ 
438538670849538432 & 41.8300422 & 48.2166719 & 16.8646& 0.00210 & 210 & 8.90&15.9012&0.00582 & 21&0.82 & 17.7918&0.00582&21 & 0.57 & 0 \\ 
444000804098995328 & 55.1883997 & 53.0607704 & 17.4842& 0.00264 & 233 & 8.67&16.4382&0.01028 & 27&1.97 & 18.5849&0.01028&28 & 0.08 & 0 \\ 
454684140113693824 & 39.4934919 & 56.0635963 & 18.9622& 0.00916 & 288 & 20.73&18.2259&0.02693 & 30&0.72 & 19.5495&0.02693&29 & 2.36 & 0 \\ 
461998465116203648 & 47.6646909 & 59.0071831 & 18.5891& 0.00877 & 549 & 51.24&17.3002&0.01131 & 53&0.82 & 19.4288&0.01131&54 & 1.05 & 0 \\ 
462961469803934848 & 47.5753587 & 60.2087274 & 15.6061& 0.00075 & 422 & 3.59&14.7368&0.00239 & 43&1.94 & 16.3636&0.00239&44 & 0.20 & 0 \\ 
486794616988335872 & 53.0359731 & 62.2067988 & 18.1025& 0.00360 & 296 & 9.72&16.9914&0.01141 & 30&1.50 & 19.1999&0.01141&28 & 0.91 & 0 \\ 
810864266136126848 & 140.9142416 & 36.9725754 & 16.0517& 0.00489 & 244 & 40.04&15.7567&0.01417 & 25&10.33 & 16.1887&0.01417&26 & 9.00 & 0 \\ 

... & ... &... &... &... &... &... &... &... &... &... &... &... &... &... &... 

\enddata
\label{tab:CandidateVariablesTable}
\tablecomments{Col. (1) Gaia DR2 source ID. Cols. (2) \& (3) Right Ascension and Declination. Col. (4) G-band mean magnitude. Col. (5) G-band magnitude error. Col. (6) Number of G-band photometric observations. Col. (7) Number of $\sigma$ G-band magnitude error is off baseline curve. Col. (8) RP-band mean magnitude. Col. (9) RP-band magnitude error. Col. (10) Number of RP-band photometric observations. Col. (11) Number of $\sigma$ RP-band magnitude error is off baseline curve. Col.(12) 
BP-band mean magnitude. Col. (13) BP-band magnitude error. Col. (14) Number of BP-band photometric observations. Col. (15) Number of $\sigma$ BP-band magnitude error is off baseline curve. Col. (16) has a value of 0 if star does not have a nearby neighbor meeting the criteria listed in Section \ref{sec:FalsePositives}, and a value of 1 if star does have nearby neighbor.}
\end{deluxetable}
\clearpage
\end{turnpage}

\begin{comment}
\begin{figure}
\begin{tabular}{cccc}
\subfloat[]{\includegraphics[width = 1.5in]{SampleVariable1.png}} &
\subfloat[]{\includegraphics[width = 1.5in]{SampleVariable2.png}} &
\subfloat[]{\includegraphics[width = 1.5in]{SampleVariable3.png}} &
\subfloat[]{\includegraphics[width = 1.5in]{SampleVariable4.png}}\\
\subfloat[]{\includegraphics[width = 1.5in]{SampleVariable5.png}} &
\subfloat[]{\includegraphics[width = 1.5in]{SampleVariable6.png}} &
\subfloat[]{\includegraphics[width = 1.5in]{SampleVariable7.png}} &
\subfloat[]{\includegraphics[width = 1.5in]{SampleVariable8.png}}\\
\subfloat[]{\includegraphics[width = 1.5in]{SampleVariable9.png}} &
\subfloat[]{\includegraphics[width = 1.5in]{SampleVariable10.png}} &
\subfloat[]{\includegraphics[width = 1.5in]{SampleVariable11.png}} &
\subfloat[]{\includegraphics[width = 1.5in]{SampleVariable12.png}}\\
\subfloat[]{\includegraphics[width = 1.5in]{SampleVariable13.png}} &
\subfloat[]{\includegraphics[width = 1.5in]{SampleVariable19.png}} &
\subfloat[]{\includegraphics[width = 1.5in]{SampleVariable17.png}} &
\subfloat[]{\includegraphics[width = 1.5in]{SampleVariable16.png}}
\end{tabular}
\caption{Example light curves of 16 candidate variables}
\label{fig:16CandidateVariables}
\end{figure}
\end{comment}

\end{document}